\documentclass[12pt]{article}

\usepackage{makeidx}
\usepackage{graphicx}
\usepackage{wrapfig}
\newcommand{\bi}{\bibitem}
\newcommand{\be}{\begin{eqnarray}}
\newcommand{\ee}{\end{eqnarray}}

\usepackage[]{caption}
\def\fun#1#2{\lower3.6pt\vbox{\baselineskip0pt\lineskip.9pt
\ialign{$\mathsurround=0pt#1\hfil##\hfil$\crcr#2\crcr\sim\crcr}}}

\begin{document}

\begin{titlepage}
\title{ CPT, Lorentz invariance, mass differences, and charge non-conservation}
\author{A.D. Dolgov$^{a,b,c,d}$, V.A. Novikov$^{a,b}$}

\maketitle
\begin{center}
$^a$ Novosibirsk State University, Novosibirsk, 630090, Russia\\
$^b$ Institute of Theoretical and Experimental Physics, Moscow, 113259, Russia \\
$^{c}$Dipartimento di Fisica, Universit\`a degli Studi di Ferrara, I-44100 Ferrara, Italy \\
$^{d}$Istituto Nazionale di Fisica Nucleare, Sezione di Ferrara,
I-44100 Ferrara, Italy
\end{center}




\begin{abstract}
A non-local field theory which breaks discrete symmetries, including C, P, CP, and CPT,
but preserves Lorentz symmetry, is presented. We demonstrate that at one-loop level
the masses for particle and antiparticle remain equal due to Lorentz symmetry only. An
inequality of masses implies breaking of the Lorentz invariance and non-conservation of
the usually conserved charges.

\end{abstract}

\end{titlepage}

\section{Introduction}

The interplay of Lorentz symmetry and CPT symmetry was considered
in the literature for decades. The issue attracted  an additional
interest recently due to a CPT-violating scenario in neutrino
physics with different mass spectrum of neutrinos and
antineutrinos~\cite{1}. Theoretical frameworks of CPT breaking in
quantum field theories, in fact in string theories, and detailed
phenomenology of oscillating neutrinos with different masses of
$\nu$ and $\bar \nu$ was further studied in papers~\cite{2}.

On the other hand, it was argued in ref.~\cite{3}
that violation of CPT automatically leads to violation of the
Lorentz symmetry \cite{3}. This might allow for some more freedom in phenomenology
of neutrino oscillations.

Very recently this conclusion was revisited in our paper \cite{4}.
We demonstrated that field theories with different masses for
particle and antiparticle are extremely pathological ones and
can't be treated as healthy quantum field theories. Instead we
constructed a class of slightly non-local Lorentz invariant field
theories with the explicit breakdown of CPT symmetry and with the
same masses for particle and  antiparticle.

An example of such theory is a non-local QED with  the Lagrangian
${\cal L} = {\cal L}_0 + {\cal L}_{n.l.}$, where ${\cal L}_0$ is the usual QED
Lagrangian:
\begin{equation}
{\cal L}_0 = -\frac{1}{4}F_{\mu\nu}^{2}(x) + \bar\psi(x)[i\hat\partial - e\hat A(x) -m] \psi(x) \;\; , \label{1}
\end{equation}
and ${\cal L}_{n.l}$ is a small non-local addition:
\begin{equation}
{\cal L}_{n.l.}(x) = g\int dy \bar\psi(x) \gamma_\mu \psi(x)
A_\mu(y) K(x-y) \;\; , \label{2}
\end{equation}
Here $F_{\mu\nu}(x) = \partial_\mu A_\nu(x) - \partial_\nu A_\mu(x)$ is
the electromagnetic field strength tensor, $A_\mu(x)$ is the four-potential, and  $\psi(x)$
is the Dirac field for electrons.

Non-local form-factor $K(x-y)$ is chosen in such a way that it explicitly breaks  $T$-invariance, e.g.
\begin{equation}
K(x-y) = \theta(x_0 - y_0) \theta[(x-y)^2] e^{-(x-y)^2/l^2} \;\; ,
\label{3}
\end{equation}
where $l$ is a scale of the non-locality and the Heaviside
functions $\theta(x_0 - y_0)\theta[(x-y)^2]$ are equal to the
unity for the future light-cone and are identically zero for the
past light-cone.

Non-local interaction, eq.~(2), breaks T-invariance, preserves C-
and P-invariance and, as a result, breaks CPT-invariance. This
construction demonstrates that CPT-symmetry can be broken in
Lorentz-invariant non-local field theory! The masses of an
electron, $m$, and of a positron, $\tilde m$, remain identical to
each other in this theory despite breaking of CPT-symmetry. The
evident reason is  that the interaction ${\cal L}_{n.l.}(x)$ is
C-invariant and its exact C-symmetry preserves the identity of
masses and anti-masses.

In this note we would like to study further the relation between
mass difference for a particle and an antiparticle and
CPT-symmetry. We start from the standard local free field theory
of electrons with the usual dispersion relation between energy and
momentum:
\begin{equation}
p_\mu^2 = p_0^2 -{\bf p}^2 = m^2 = \tilde m^2 \label{4}
\end{equation}
and introduce a non-local interaction that breaks the whole set of
discrete symmetries, i.e. C, P, CP, T, and CPT. So there is no
discrete symmetry which preserves equality of $m$ to $\tilde m$ in this case. Hence in
principle the interaction can shift $m$ from $\tilde m$. But an explicit
one-loop calculation demonstrates that this is not true. So we
conclude that it is Lorentz-symmetry that keeps the identity
\begin{equation}
m=\tilde m \;\; . \label{5}
\end{equation}

This conclusion invalidates the experimental evidence for
CPT-symmetry based on the equality of masses of particles and
antiparticles. CPT may be strongly broken in a Lorentz invariant
way and in such a case the masses must be equal. Another way
around, if we assume that the masses are different, then Lorenz
invariance must be broken. Lorentz and CPT violating theories
would lead not only to mass difference of particles and
antiparticles but to much more striking phenomena such as
violation of gauge invariance, current non-conservation, and even
to a breaking of the usual equilibrium statistics (for the latter
see ref.~\cite{ad-cpt}).

\section{C, CP and CPT violating QFT}

To formulate a model we start with the standard QED Lagrangian:
\begin{equation}
{\cal L}_0 = -\frac{1}{4}F_{\mu\nu}(x) F_{\mu\nu}(x)
+\bar\psi(x)[i\hat\partial -e\hat A(x)-m]\psi(x) \;\; , \label{6}
\end{equation}
and add the interaction of a photon, $A_\mu$, with an axial
current
\begin{equation}
{\cal L}_1 = g_1\bar\psi(x) \gamma_\mu \gamma_5 \psi(x) A_\mu(x)
\label{7}
\end{equation}
and with the electric dipole moment of an electron
\begin{equation}
{\cal L}_2 = g_2\bar\psi(x)\sigma_{\mu\nu}\gamma_5
\psi(x)F_{\mu\nu}(x) \;\; . \label{8}
\end{equation}

The first interaction, ${\cal L}_1$, breaks C and P-symmetry and
conserves CP-symmetry. The second interaction breaks P- and
CP-symmetry. Still the sum of Lagrangians
\begin{equation}
{\cal L} = {\cal L}_0 +{\cal L}_1 +{\cal L}_2 \label{9}
\end{equation}
preserves CPT-symmetry. To break the CPT we  modify the
interaction ${\cal L}_1$ to a non-local one $\tilde{\cal L}_1$:
\begin{equation}
{\cal L}_1\to \tilde{\cal L}_1(x) = \int dy g_1
\bar\psi(x)\gamma_\mu \gamma_5 \psi(x) K(x-y) A_\mu(y) \;\; .
\label{10}
\end{equation}
With this modification the model
\begin{equation}
{\cal L} = {\cal L}_0 +\bar{\cal L}_1+{\cal L}_2 \label{11}
\end{equation}
breaks all discrete symmetries.

\section{One-loop calculation}
In general to calculate high order perturbative contributions of a
non-local interaction into $S$-matrix one has to modify the Dyson
formulae for $S$-matrix with $T$-ordered exponential
\begin{equation}
S=T\left\{{\rm exp}\left( i {\int d^4 x}  {\cal L}_{int}\right) \right\} \label{12}
\end{equation}
and the whole Feynman diagram techniques.

But in the first order in the non-local interaction one can work
with the usual Feynman rules in the coordinate space. The only
difference is that one of the vertices becomes non-local.

\section{Mass and wave function renormalization for
particle and antiparticle}

We start with the standard free field theory for an electron, i.e.
\begin{equation}
{\cal L} = \bar\psi[i\hat\partial -m]\psi \label{13}
\end{equation}
that fixes the usual dispersion law
\begin{equation}
p^2 = p_0^2 -{\rm\bf p}^2 = m^2 \;\; . \label{14}
\end{equation}

The self-energy operator, $\Sigma(p)$, contributes both to the mass
renormalization and to the wave function renormalization. In general one-loop effective
Lagrangin can be written in the form:
\begin{equation}
{\cal L}_{eff}^{(1)} = \bar\psi[i(A\gamma_\mu + B\gamma_\mu
\gamma_5)\partial_\mu -(m_1 + im_2 \gamma_5)]\psi \;\; .
\label{15}
\end{equation}

It is useful to rewrite the same one-loop effective Lagrangian in
terms of the field for antiparticle $\psi_c$:
\begin{equation}
\psi_c = (-i)[\bar\psi\gamma^0 \gamma^2]^T \;\; , \label{16}
\end{equation}

\begin{equation}
{\cal L}_{eff}^{(1)} = \bar\psi_c[i(A\gamma_5 -B\gamma_\mu
\gamma_5)\partial_\mu -(m_1 + im_2 \gamma_5)]\psi_c \;\; .
\label{17}
\end{equation}

We see that the mass term is the same for $\psi$ and for $\psi_c$, but
the wave function renormalization is different: the coefficient in front of
the pseudovector changes its sign. This change is unobservable since
one can remove $B\gamma_\mu \gamma_5$ and $im_2\gamma_5$ terms by
redefining of variables. Indeed
\begin{equation}
\bar\psi(A+B\gamma_5)\gamma_\mu\psi \equiv
\bar\psi^\prime\sqrt{A^2 +B^2}\gamma_\mu\psi^\prime \;\; ,
\label{18}
\end{equation}
where
\begin{equation}
\psi = (\cosh \alpha + i\gamma_5 \, \sinh \alpha)\psi^\prime \;\; ,
\label{19}
\end{equation}
\begin{equation}
\tanh2 \alpha = B/A \;\; , \label{20}
\end{equation}
and
\begin{equation}
\bar\psi(m_1 + i\gamma_5 m_2)\psi \equiv \sqrt{m_1^2 +
m_2^2}\bar\psi^\prime \psi^\prime \;\; , \label{21}
\end{equation}
where
\begin{equation}
\psi = {\rm exp}{(i\gamma_5\beta)} \psi^\prime \;\; ,
\label{22}
\end{equation}

\begin{equation}
\tan2\beta = m_2/m_1 \;\; . \label{23}
\end{equation}

This simple observation is sufficient to conclude that technically there is no
possibility to write one-loop corrections that produce different
contributions for particle and antiparticle.
Still it is instructive to check directly that the difference is zero.

\section{Explicit one-loop calculation}

We are looking for a  one-loop contribution into self-energy
operator $\Sigma(p)$ that breaks C, CP, and CPT  symmetries and
that changes the chirality of the fermion line. It is clear that
this contribution potentially can be different (opposite in sign)
for particle $\psi$ and antiparticle $\psi_c$.

To construct such  contribution we need both anomalous
interactions  $\tilde{\cal L}_1$ and ${\cal L}_2$. Indeed
interaction ${\cal L}_2$  changes chirality and breaks CP
symmetry, while non-local interaction $\tilde{\cal L}_1$ breaks C
and CPT and leaves the chirality unchanged. In combination they
break all discrete symmetries and change chirality. There are two
diagrams that are proportional to $g_1 g_2$ (see Fig. 1).

\begin{figure}
 \centering
 \includegraphics{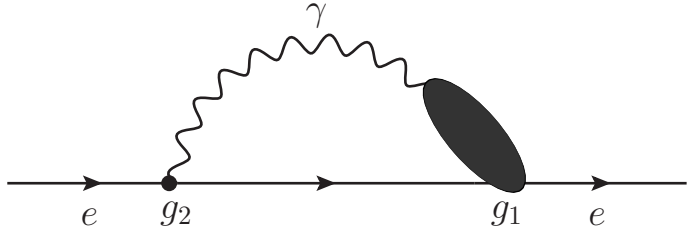}
 \caption{The diagram contributing to the mass difference of electron and positron. The blob represents a non-local
form-factor.}
\end{figure}

We will calculate these diagrams in two steps. The first step is a
pure algebraic one. Self-energy $\Sigma(p)$  is 4$\times$4 matrix
that was constructed from a product of three  other 4$\times$4
matrices, i.e. two vertices and one fermion propagator. Notice
that any 4$\times$4  matrix can be decomposed as a sum over
complete set of 16 Dirac matrices. In this decomposition of
$\Sigma(p)$ we need terms that are odd in $C$ and changes
chirality. Fortunately there is only one Dirac matrix with these
properties. That is $\sigma_{\mu\nu}$. So
\begin{equation}
\Sigma(p) = \sigma_{\mu\nu} I_{\mu\nu} \;\;(p) , \label{24}
\end{equation}
where $I_{\mu\nu}$ represents Feynman (divergent) integral. We
could obtain eq.~(24) after some long explicit algebraic
transformation, but the net result is determined by the symmetry
only.

The second step is the calculation of Feynman integrals. Again
fortunately we do not need actual calculations. Indeed due to the
Lorentz symmetry of the theory this  $I_{\mu\nu}$ should be a
tensor that depends only on the momentum of fermion line $p$. The
general form for $I_{\mu\nu}$ is
\begin{equation}
I_{\mu\nu} = Ag_{\mu\nu} + Bp_\mu p_\nu \;\; . \label{25}
\end{equation}

As a result we get
\begin{equation}
\Sigma(p) = \sigma_{\mu\nu}I_{\mu\nu} \equiv 0 \;\;  \label{26}
\end{equation}
and we conclude that the one-loop contribution into possible mass
difference is identically zero.\footnote{Recently our former
collaborators published a paper where they demonstrated that for a
particle with a non-standard dispersion law the quantity which
they define as mass can be different for particle and antiparticle
\cite{4}.}

\section{CPT and charge non-conservation}

There is  widely spread habit to parametrize $CPT$ violation by
attributing  different masses to particle and antiparticle. This
tradition is traced to an old time of the first observation of
$K-{\bar K}$-mesons oscillation.

For  $K$-mesons with a given momenta ${\bf q}$ the theory of
oscillation is equivalent to a non-hermitian Quantum Mechanics
(QM) with two degrees of freedom. Diagonal elements of $2 \times 2$
Hamiltonian matrix represent masses for particle and antiparticle.
Their unequality breaks CPT-symmetry. Experimental bounds on mass
difference are considered as bounds on the CPT-symmetry violation parameters.
Such  strategy has no explicit loop-holes and is still used for parametrization of
CPT-symmetry violation in  $D$ and $B$  meson oscillations.

Quantum Field Theory (QFT) deals not with one mode for a given
momenta but rather with an infinite sum over all  momenta. The set
of plane waves with all possible momenta for particle and
antiparticle is a complete set of orthogonal modes and an
arbitrary field operator can be decomposed over this set.

Naive generalization of  CPT-conserving QFT to CPT-violating QFT
was to attribute  different masses for particle and antiparticle
\cite{1,2}). Say for a complex scalar field they use the infinite
sum \cite{1,2}
\begin{equation}
\phi(x) = \sum_{\rm\bf q}\left\{a({\rm\bf q})\frac{1}{\sqrt{2E}}e^{-i(Et-{\bf qx})}  + b^+({\rm\bf q})
\frac{1}{\sqrt{2\tilde E}} e^{i({\tilde E}t-\rm\bf qx)}\right\}, \;\;
\label{27}\
\end{equation}
where ($ a({\bf q}), a^+({\bf q})$), ($ b({\bf q})$, $ b^+({\bf
q})$) are annihilation and creation operators, and ($m, E$) and
$({\tilde m}, {\tilde E})$ are masses and energies of particle and
antiparticle respectively.

Greenberg \cite{3} found that this construction runs into trouble. The dynamic
of fields determined according to  eq.~(27)  cannot be a  Lorentz-invariant one.

We'd like to notice that for charged particles (say for electrons
and positrons) similar generalization of the field theory breaks
not only the Lorentz symmetry but the electric charge conservation
as well. The reason is very simple. For the standard QED the
operator of electric charge ${\hat Q(t)}$ can be written in the
form
\begin{equation}
{\hat Q(t)}  =  \sum_{\rm\bf q}\left\{a^+({\rm\bf q}) a({\rm\bf q}) -
b^+({\rm\bf q}) b({\rm\bf q})\right\} \;\; . \label{28}
\end{equation}
Operator $\hat Q(t)$ is a diagonal one, i.e. there are no mixed terms with different
momenta. The modes with different momenta are orthogonal to each other and disappear
after integration over space. This is a technical explanation  why one can construct
a time-independent operator.

If one shifts the mass of electron from the mass of positron the
situation drastically changes. For electron the modes  with
different momenta are still orthogonal to each other. The same is
true for the modes of positron, they are also orthogonal among
themselves. But there is no reason for wave function of electron
with mass $m$ be orthogonal to wave functions for positron with
mass $\tilde m$. As a result one obtaines
\begin{equation}
Q(t) = \sum_{\rm\bf q}\left\{a^+({\rm\bf q}) a({\rm\bf q}) -
b^+({\rm\bf q}) b({\rm\bf q})\right\} + \nonumber \  C \sum_{\rm\bf q} \frac{(E-\tilde E)}{\sqrt{4E\tilde E}}
\left[b({\rm\bf q})a(-{\rm\bf q})e^{-i(E+\tilde E)t} + {\rm
h.c.}\right] \,, \label{29}
\end{equation}
where  constant $C$ depends on the sorts of particles and on the definition of the charge.

We can conclude from this equation that non-conservation of charge
exhibits itself only in annihilation processes but not in the
scattering processes. So there is no immediate problem with the
Coulomb law. Nevertheless non-conservation of this type is also
absolutely excluded by the experiment. In a case of
charge-nonconservation annihilation of particle and antiparticle
with a creation of the infinite number of soft massless photons
creates a terrible infrared problem. Infrared catastrophe can not
be avoided by usual summation over infrared photons. On the other
hand, as is argued in ref.~\cite{OVZ},  the electron decay might
be exponentially suppressed due to vanishing of the corresponding
formfactor created by virtual longitudinal photons.

Similar arguments lead to the conclusion that conservation  of
energy cannot  survive as well in a theory with different masses
of particles and antiparticles.

\section{Conclusion}

We have shown that in the framework of a Lorentz invariant field theory it is
impossible to have different masses of particles and antiparticles, even if CPT (together
with C and P) invariance is broken. On the other hand, unequal masses of particles and
antiparticles imply breaking of the Lorentz invariance. Moreover, in such theories charge
and energy conservation seem to be broken as well.

\section{Acknowledgements}

This work was supported by the Grant of Government of Russian Federation
(11.G34.31.0047), by NSh-3172.2012.2,  and by the Grant RFBR 11-02-00441.

\end{document}